\documentclass[a4paper,11pt]{article}
\usepackage{a4,amsmath,amsfonts,epsfig,cite}

\newcommand{\blockA}[3]{\ensuremath{
     \setlength{\unitlength}{1mm}
     \begin{picture}(10,15)(0,0)
          \put( 0, 4){\line(1,0){10}}
          \put(10, 4){\line(0,1){10}}
          \put( 2, 5){\makebox(0,0)[lb]{\small $#1$}}
          \put(11,13){\makebox(0,0)[lt]{\small $#2$}}
          \put(10, 3){\makebox(0,0)[ct]{\small $#3$}}
     \end{picture}}}

\newcommand{\blockB}[1]{\ensuremath{
     \setlength{\unitlength}{1mm}
     \begin{picture}(10,15)(0,0)
          \put( 0, 4){\line(1,0){10}}
          \put( 8, 5){\makebox(0,0)[rb]{\small $#1$}}
     \end{picture}}}

\newcommand{\blockC}[3]{\ensuremath{
     \setlength{\unitlength}{1mm}
     \begin{picture}(10,15)(0,0)
          \put( 0, 4){\line(1,0){10}}
          \put(10, 4){\line(0,1){10}}
          \put( 5, 5){\makebox(0,0)[cb]{\small $#1$}}
          \put(11,13){\makebox(0,0)[lt]{\small $#2$}}
          \put(10, 3){\makebox(0,0)[ct]{\small $#3$}}
     \end{picture}}}

\newcommand{\blockD}[3]{\ensuremath{
     \setlength{\unitlength}{1mm}
     \begin{picture}(8,15)(0,0)
          \put( 0, 4){\line(1,0){8}}
          \put( 8, 4){\line(0,1){10}}
          \put( 1, 5){\makebox(0,0)[cb]{\small $#1$}}
          \put( 9,13){\makebox(0,0)[lt]{\small $#2$}}
          \put( 8, 3){\makebox(0,0)[ct]{\small $#3$}}
     \end{picture}}}

\newcommand{\blockE}[0]{\ensuremath{
     \setlength{\unitlength}{1mm}
     \begin{picture}(3,15)(0,0)
          \put( 0, 4){\line(1,0){3}}
     \end{picture}}}

\newcommand{\blockUp}[1]{\ensuremath{
     \setlength{\unitlength}{1mm}
     \put( 0,10){#1} }}

\newcommand{\cbB}[7]{\blockA{#1}{#2}{#6}\blockC{#5}{#3}{#7}\blockB{#4}}
\newcommand{\cbC}[7]{\blockA{#1}{}{#6}%
     \blockUp{\blockD{#5}{#2}{#7}\blockB{#3}}%
     \blockE\blockB{#4}}

\newcommand{\bL}[1]{\raisebox{-3mm}{#1}}

\newcommand{\F}[6]{\ensuremath{
     \textsf{\large F}_{\raisebox{-1pt}{\scriptsize\!\!#5#6}}\!\left[
     {#2 \atop #1}\;{#3 \atop #4}\right]}}

\newcommand{\Bmat}[7]{\ensuremath{
     \textsf{\large B}^{(#1)}_{\raisebox{-1pt}{\scriptsize\!$#6#7$}}\!\left[
     {#3 \atop #2}\;{#4 \atop #5}\right]}}

\newcommand{\ffbnd}[8]{\ensuremath{
     \raisebox{-2mm}{
     \setlength{\unitlength}{0.7mm}
     \begin{picture}(50,10)(0,0)
          \put( 0, 5){\line(1,0){9}}
          \put(11, 5){\line(1,0){8}}
          \put(21, 5){\line(1,0){8}}
          \put(31, 5){\line(1,0){8}}
          \put(41, 5){\line(1,0){9}}
          \put(10, 5){\qbezier(-1,0)(-1,1)(0,1)\qbezier(0,1)(1,1)(1,0)}
          \put(20, 5){\qbezier(-1,0)(-1,1)(0,1)\qbezier(0,1)(1,1)(1,0)}
          \put(30, 5){\qbezier(-1,0)(-1,1)(0,1)\qbezier(0,1)(1,1)(1,0)}
          \put(40, 5){\qbezier(-1,0)(-1,1)(0,1)\qbezier(0,1)(1,1)(1,0)}
          \put( 5, 6){\makebox(0,0)[cb]{\small #1}}
          \put(15, 6){\makebox(0,0)[cb]{\small #2}}
          \put(25, 6){\makebox(0,0)[cb]{\small #3}}
          \put(35, 6){\makebox(0,0)[cb]{\small #4}}
          \put(45, 6){\makebox(0,0)[cb]{\small #1}}
          \put(10, 4){\makebox(0,0)[ct]{\small #5}}
          \put(20, 4){\makebox(0,0)[ct]{\small #6}}
          \put(30, 4){\makebox(0,0)[ct]{\small #7}}
          \put(40, 4){\makebox(0,0)[ct]{\small #8}}
     \end{picture}}}}

\newcommand{\hhcircle}[1]{\ensuremath{
     \setlength{\unitlength}{#1}
     \begin{picture}(2,2)(0,0)
          \put(1,1){\qbezier(-1, 0)(-1, 1)( 0, 1)}
          \put(1,1){\qbezier( 0, 1)( 1, 1)( 1, 0)}
          \put(1,1){\qbezier( 1, 0)( 1,-1)( 0,-1)}
          \put(1,1){\qbezier( 0,-1)(-1,-1)(-1, 0)}
     \end{picture}}}

\newcommand{\hhpicUHP}{\ensuremath{
     \raisebox{-1mm}{
     \setlength{\unitlength}{0.5mm}
     \begin{picture}(22,8)(0,0)
          \put( 0, 2){\line(1,0){9}}
          \put(11, 2){\line(1,0){9}}
          \put(10, 2){\qbezier(-1,0)(-1,1)(0,1)\qbezier(0,1)(1,1)(1,0)}
          \put( 5, 3){\makebox(0,0)[cb]{\small a}}
          \put(15, 3){\makebox(0,0)[cb]{\small b}}
     \end{picture}}}}

\newcommand{\hhpicRING}{\setlength{\unitlength}{1mm}
     \raisebox{-2mm}{
     \begin{picture}(8,6)(0,0)
          \put(1.5,1.5){\hhcircle{1.5mm}}
          \put(0,0){\hhcircle{3mm}}
          \put(2.5,3){\makebox(0,0)[lc]{\small a}}
          \put(8,3){\makebox(0,0)[rc]{\small b}}
     \end{picture}}}

\newcommand{\B}[3]{\ensuremath{\,{^{(\!#1\!)}}{\!B_#2}^{#3}\,}}

\newcommand{\bc}[6]{\ensuremath{\,C^{(\!#1\hspace{-0.7pt}#2\hspace{-0.7pt}#3\!)#6}_{\;#4#5}\,}}
\newcommand{\bnpt}[2]{\ensuremath{\langle{#1}\rangle^{\!(\!#2\!)}}}
\newcommand{\bp}[3]{\ensuremath{\,\psi^{(\!#1\hspace{-0.7pt}#2\!)}_{#3}}}
\newcommand{\C}[3]{\ensuremath{\,{C_{{#1}{#2}}}^{#3}\,}}
\newcommand{\D}{\ensuremath{\Delta}}
\newcommand{\G}[1]{\ensuremath{\Gamma\left({\textstyle #1}\right)}}
\newcommand{\h}[2]{\ensuremath{h_{(#1,#2)}}}

\newcommand{\is}[1]{\ensuremath{|#1\rangle}}
\newcommand{\ishin}[1]{\ensuremath{|#1\rangle\!\rangle}}
\renewcommand{\l}{\ensuremath{\ell}}

\newcommand{\M}[2]{\ensuremath{\mathrm{M}({#1},{#2})}}
\newcommand{\N}[3]{\ensuremath{{\mathrm{N}_{#1#2}}^{#3}}}
\newcommand{\npt}[1]{\ensuremath{\langle{#1}\rangle}}
\renewcommand{\O}{\ensuremath{\Omega}}
\newcommand{\os}[1]{\ensuremath{\langle{#1}|}}
\newcommand{\ph}[1]{\ensuremath{\phi_{#1}}}
\newcommand{\phs}[2]{\ensuremath{\phi_{({#1},{#2})}}}
\newcommand{\pp}{\ensuremath{p'}}

\renewcommand{\S}[2]{\ensuremath{{S_{#1}}^{#2}}}

\newcommand{\simu}[2]{\ensuremath{\underset{#1\rightarrow #2}{\sim}}}

\begingroup\makeatletter\ifx\SetFigFont\undefined
\def\x#1#2#3#4#5#6#7\relax{\def\x{#1#2#3#4#5#6}}%
\expandafter\x\fmtname xxxxxx\relax \def\y{splain}%
\ifx\x\y   
\gdef\SetFigFont#1#2#3{%
  \ifnum #1<17\tiny\else \ifnum #1<20\small\else
  \ifnum #1<24\normalsize\else \ifnum #1<29\large\else
  \ifnum #1<34\Large\else \ifnum #1<41\LARGE\else
     \huge\fi\fi\fi\fi\fi\fi
  \csname #3\endcsname}%
\else
\gdef\SetFigFont#1#2#3{\begingroup
  \count@#1\relax \ifnum 25<\count@\count@25\fi
  \def\x{\endgroup\@setsize\SetFigFont{#2pt}}%
  \expandafter\x
    \csname \romannumeral\the\count@ pt\expandafter\endcsname
    \csname @\romannumeral\the\count@ pt\endcsname
  \csname #3\endcsname}%
\fi
\fi\endgroup

\newcommand{\figal}{\begin{picture}(0,0)%
\epsfig{file=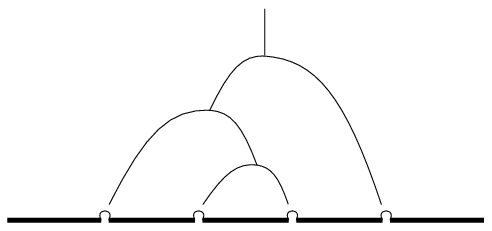}%
\end{picture}%
\setlength{\unitlength}{0.00041700in}%
\begin{picture}(4663,2820)(260,-2515)
\put(754,-2014){\makebox(0,0)[b]{\smash{\SetFigFont{9}{10.8}{rm}$a$}}}
\put(1729,-2014){\makebox(0,0)[b]{\smash{\SetFigFont{9}{10.8}{rm}$b$}}}
\put(2629,-2014){\makebox(0,0)[b]{\smash{\SetFigFont{9}{10.8}{rm}$c$}}}
\put(3529,-2014){\makebox(0,0)[b]{\smash{\SetFigFont{9}{10.8}{rm}$d$}}}
\put(4429,-2014){\makebox(0,0)[b]{\smash{\SetFigFont{9}{10.8}{rm}$a$}}}
\put(1276,-2461){\makebox(0,0)[b]{\smash{\SetFigFont{9}{10.8}{rm}$i$}}}
\put(2176,-2461){\makebox(0,0)[b]{\smash{\SetFigFont{9}{10.8}{rm}$j$}}}
\put(3076,-2461){\makebox(0,0)[b]{\smash{\SetFigFont{9}{10.8}{rm}$k$}}}
\put(2926, 89){\makebox(0,0)[b]{\smash{\SetFigFont{9}{10.8}{rm}$\bnpt 1a$}}}
\put(2776,-1561){\makebox(0,0)[lb]{\smash{\SetFigFont{9}{10.8}{rm}$\bc bcdjkq$}}}
\put(3976,-2461){\makebox(0,0)[b]{\smash{\SetFigFont{9}{10.8}{rm}$\l$}}}
\put(2251,-961){\makebox(0,0)[rb]{\smash{\SetFigFont{9}{10.8}{rm}$\bc abdiq\l$}}}
\put(2926,-511){\makebox(0,0)[lb]{\smash{\SetFigFont{9}{10.8}{rm}$\bc ada\l\l1$}}}
\end{picture}}
\newcommand{\figar}{\begin{picture}(0,0)%
\epsfig{file=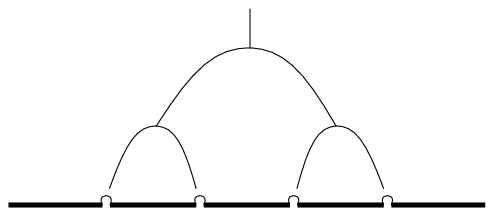}%
\end{picture}%
\setlength{\unitlength}{0.00041700in}%
\begin{picture}(4754,2751)(172,-2671)
\put(757,-2092){\makebox(0,0)[b]{\smash{\SetFigFont{9}{10.8}{rm}$a$}}}
\put(1732,-2092){\makebox(0,0)[b]{\smash{\SetFigFont{9}{10.8}{rm}$b$}}}
\put(2632,-2092){\makebox(0,0)[b]{\smash{\SetFigFont{9}{10.8}{rm}$c$}}}
\put(3532,-2092){\makebox(0,0)[b]{\smash{\SetFigFont{9}{10.8}{rm}$d$}}}
\put(1276,-2536){\makebox(0,0)[b]{\smash{\SetFigFont{9}{10.8}{rm}$i$}}}
\put(3976,-2536){\makebox(0,0)[b]{\smash{\SetFigFont{9}{10.8}{rm}$\l$}}}
\put(1951,-1411){\makebox(0,0)[lb]{\smash{\SetFigFont{9}{10.8}{rm}$\bc abcijp$}}}
\put(4432,-2092){\makebox(0,0)[b]{\smash{\SetFigFont{9}{10.8}{rm}$a$}}}
\put(3076,-2536){\makebox(0,0)[b]{\smash{\SetFigFont{9}{10.8}{rm}$k$}}}
\put(2176,-2536){\makebox(0,0)[b]{\smash{\SetFigFont{9}{10.8}{rm}$j$}}}
\put(3676,-1411){\makebox(0,0)[lb]{\smash{\SetFigFont{9}{10.8}{rm}$\bc cdak\l p$}}}
\put(2701,-586){\makebox(0,0)[lb]{\smash{\SetFigFont{9}{10.8}{rm}$\bc acapp1$}}}
\put(2851,-136){\makebox(0,0)[b]{\smash{\SetFigFont{9}{10.8}{rm}$\bnpt 1a$}}}
\end{picture}}

\newcommand{\figbl}{\begin{picture}(0,0)%
\epsfig{file=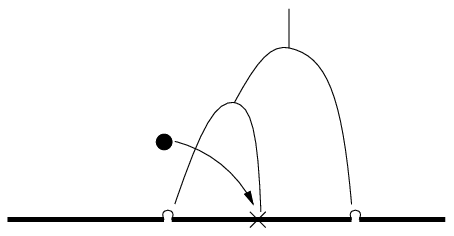}%
\end{picture}%
\setlength{\unitlength}{0.00041700in}%
\begin{picture}(4500,2667)(1531,-2362)
\put(5447,-1937){\makebox(0,0)[b]{\smash{\SetFigFont{9}{10.8}{rm}$a$}}}
\put(3180,-2308){\makebox(0,0)[b]{\smash{\SetFigFont{9}{10.8}{rm}$p$}}}
\put(5023,-2306){\makebox(0,0)[b]{\smash{\SetFigFont{9}{10.8}{rm}$q$}}}
\put(2580,-1933){\makebox(0,0)[b]{\smash{\SetFigFont{9}{10.8}{rm}$a$}}}
\put(3480,-1933){\makebox(0,0)[b]{\smash{\SetFigFont{9}{10.8}{rm}$b$}}}
\put(2955,-1183){\makebox(0,0)[b]{\smash{\SetFigFont{9}{10.8}{rm}$i$}}}
\put(4080,-1783){\makebox(0,0)[lb]{\smash{\SetFigFont{9}{10.8}{rm}$\B bi\l$}}}
\put(4276,-361){\makebox(0,0)[rb]{\smash{\SetFigFont{9}{10.8}{rm}$\bc abaqq1$}}}
\put(4426, 89){\makebox(0,0)[b]{\smash{\SetFigFont{9}{10.8}{rm}$\bnpt 1a$}}}
\put(3826,-961){\makebox(0,0)[lb]{\smash{\SetFigFont{9}{10.8}{rm}$\bc abbp\l q$}}}
\end{picture}}
\newcommand{\figbr}{\begin{picture}(0,0)%
\epsfig{file=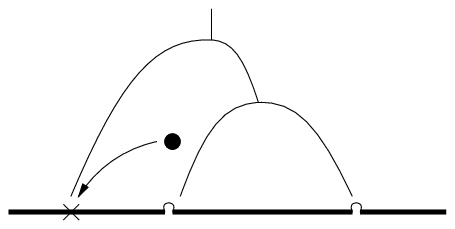}%
\end{picture}%
\setlength{\unitlength}{0.00041700in}%
\begin{picture}(4447,2671)(325,-2366)
\put(3901,-2311){\makebox(0,0)[b]{\smash{\SetFigFont{9}{10.8}{rm}$q$}}}
\put(4428,-1938){\makebox(0,0)[b]{\smash{\SetFigFont{9}{10.8}{rm}$a$}}}
\put(2177,-2312){\makebox(0,0)[b]{\smash{\SetFigFont{9}{10.8}{rm}$p$}}}
\put(1279,-1712){\makebox(0,0)[rb]{\smash{\SetFigFont{9}{10.8}{rm}$\B aik$}}}
\put(1728,-1938){\makebox(0,0)[b]{\smash{\SetFigFont{9}{10.8}{rm}$a$}}}
\put(3151,-1936){\makebox(0,0)[b]{\smash{\SetFigFont{9}{10.8}{rm}$b$}}}
\put(2626, 89){\makebox(0,0)[b]{\smash{\SetFigFont{9}{10.8}{rm}$\bnpt 1a$}}}
\put(2551,-361){\makebox(0,0)[lb]{\smash{\SetFigFont{9}{10.8}{rm}$\bc aaakk1$}}}
\put(1951,-1261){\makebox(0,0)[b]{\smash{\SetFigFont{9}{10.8}{rm}$i$}}}
\put(3001,-1486){\makebox(0,0)[b]{\smash{\SetFigFont{9}{10.8}{rm}$\bc abapqk$}}}
\end{picture}}

\newcommand{\figcl}{\begin{picture}(0,0)%
\epsfig{file=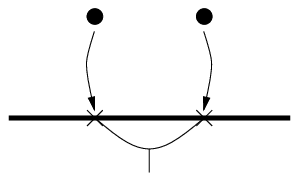}%
\end{picture}%
\setlength{\unitlength}{0.00041700in}%
\begin{picture}(2966,2366)(309,-1842)
\put(2481,308){\makebox(0,0)[b]{\smash{\SetFigFont{9}{10.8}{rm}$j$}}}
\put(1356,308){\makebox(0,0)[b]{\smash{\SetFigFont{9}{10.8}{rm}$i$}}}
\put(1876,-736){\makebox(0,0)[b]{\smash{\SetFigFont{9}{10.8}{rm}$a$}}}
\put(1279,-740){\makebox(0,0)[rb]{\smash{\SetFigFont{9}{10.8}{rm}$\B ai\l$}}}
\put(2629,-740){\makebox(0,0)[lb]{\smash{\SetFigFont{9}{10.8}{rm}$\B aj\l$}}}
\put(2102,-1788){\makebox(0,0)[b]{\smash{\SetFigFont{9}{10.8}{rm}$\bnpt 1a$}}}
\put(2102,-1338){\makebox(0,0)[lb]{\smash{\SetFigFont{9}{10.8}{rm}$\bc aaa\l\l1$}}}
\end{picture}}
\newcommand{\figcr}{\begin{picture}(0,0)%
\epsfig{file=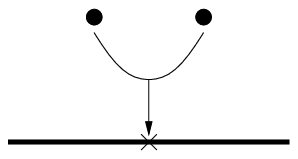}%
\end{picture}%
\setlength{\unitlength}{0.00041700in}%
\begin{picture}(3415,2564)(171,-1642)
\put(2255,310){\makebox(0,0)[b]{\smash{\SetFigFont{9}{10.8}{rm}$j$}}}
\put(1130,310){\makebox(0,0)[b]{\smash{\SetFigFont{9}{10.8}{rm}$i$}}}
\put(1879,-1489){\makebox(0,0)[b]{\smash{\SetFigFont{9}{10.8}{rm}$\bnpt 1a$}}}
\put(1729,-289){\makebox(0,0)[b]{\smash{\SetFigFont{9}{10.8}{rm}$\C ijm$}}}
\put(1654,-889){\makebox(0,0)[lb]{\smash{\SetFigFont{9}{10.8}{rm}$\B am1$}}}
\put(826,-961){\makebox(0,0)[b]{\smash{\SetFigFont{9}{10.8}{rm}$a$}}}
\end{picture}}

\begin{document}
\setcounter{footnote}{0}
\begin{titlepage}
\vskip 0.5cm
\begin{flushright}
KCL-MTH-98-59 \\
{\tt hep-th/9811178}\\
November 1998 
\end{flushright}
\vskip 1.2cm
\begin{center}
{\Large {\bf Boundary structure constants}} \\[5pt]
{\Large {\bf for the A-series Virasoro minimal models} }
\end{center}
\vskip 0.8cm
\centerline{Ingo Runkel%
\footnote{e-mail: {\tt ingo@mth.kcl.ac.uk}}
}
\vskip 0.6cm
\centerline{\sl Mathematics Department, }
\centerline{\sl King's College London, Strand, London WC2R 2LS, U.K.}
\vskip 0.9cm
\begin{abstract}
\vskip0.15cm
\noindent
We consider A-series modular invariant Virasoro minimal models on the
upper half plane. From Lewellen's sewing constraints a necessary form
of the bulk and boundary structure constants is derived. Necessary
means that any solution can be brought to the given form by a rescaling
of the fields.
All constants
are expressed essentially in terms of fusing (F-)matrix elements and the
normalisations are chosen such that they are real and
no square roots appear. It is not shown in this
paper that the given structure constants solve the sewing
constraints, however random numerical tests show no contradiction 
and agreement of the bulk structure constants
with Dotsenko and Fateev. In order to facilitate numerical calculations a
recursion relation for the F-matrices is given.
\end{abstract}
\end{titlepage}
\setcounter{footnote}{0}


\section{Introduction}
\label{j_section}

Conformal field theory (CFT) arises in the description of the
long range behaviour of statistical systems at a critical point. 
In two dimensions the symmetry algebra of the CFT becomes infinite dimensional
and places severe constraints on the theory. From the point of view of
statistical mechanics it is natural
to consider systems with boundaries instead of infinitely extended
systems. Introducing the boundary reduces the symmetry of the system,
but for certain boundary conditions an infinite subset of the original
symmetries remains unbroken. This leads to a boundary conformal field
theory (BCFT).

Virasoro minimal models (see e.g.\ \cite{BPZ84,YBk}) are an
important subclass of two dimensional CFTs. \cite{CIZ87} gives an A-D-E
type classification of the field content of these models in the
bulk. To know all correlation functions it is sufficient to know the
conformal blocks (which can be found as solutions to linear differential
equations) and the structure constants which are determined by duality
properties of the correlators. The structure constants are of interest
themselves, e.g. as an essential ingredient in TCSA
(\cite{YuZ90,DTW98}). For a CFT restricted to the upper half plane
one has to know four sets
of constants: the structure constants appearing in the operator
product expansion (OPE) of bulk fields, the structure constants of boundary
fields, how bulk fields couple to boundary fields and the one point
functions of the identity on different boundaries.

The classification of the field content and possible boundary
conditions of BCFTs was initiated by Cardy \cite{Car84,CaL91}
and further addressed e.g.\ in \cite{PSS96,FuS97,BPZ98}. Once the
field content of a BCFT is known, there is a set
of consistency conditions, the sewing constraints (see \cite{Lew92}),
which the structure constants of the theory have to obey. The main
ingredient in the constraint equations are 
the matrices that transform between different bases of conformal
blocks, which are called F-matrices (see e.g \cite{MSb90}). One can
now try to use the 
sewing constraints to determine the structure constants, and
in the case of A-type Virasoro minimal models it turns out that they
can be expressed essentially in terms of F-matrix elements. 
Explicit expressions for the bulk structure constants can be found 
in \cite{DF84} and the bulk--boundary couplings have been derived in
\cite{CaL91}. In this paper only A-invariant Virasoro minimal models
on the upper half plane are considered and for these the
remaining structure constants are derived. Care has been taken to give
the expressions in a form that 
facilitates numerical computation, e.g. no square roots appear and all
results are real.
It is not proven in this paper that the structure constants given 
solve the complete set of sewing constraints\footnote{
  In this paper only constraints arising for a CFT on the upper half
  plane are considered. For sewing constraints on non-orientable
  surfaces see \cite{FPS94}.
}, but random numerical tests showed no contradiction.

Section \ref{g_section} sets the conventions used
for conformal blocks and section \ref{k_section} gives a recursion
formula to compute the 
F-matrices. In section \ref{n_section} some results about the
classification of boundary conditions and field content are stated
and the notation for boundary fields is established. Section
\ref{d_section} gives Lewellen's sewing constraints in the diagonal
case. These are used in sections \ref{p_section} and \ref{f_section}
to compute necessary forms of all structure constants. The
implications for the
identity-1-point functions of consistency under modular transformation and
continuity on the disc are studied in section \ref{h_section}.


\section{Transformation of Conformal Blocks}
\label{g_section}

First we introduce some notation:
Let $\M{\pp}{p}$ denote an A-invariant Virasoro minimal model. It
contains only spinless primary fields $\ph i$ with conformal
weights $h_i=\bar h_i$. The weights $h_i=\h rs$ take values in the Kac-table:
\begin{align}
     \h rs &= \frac{1}{4t} \left( d^2-(t-1)^2 \right)
     \qquad \mathrm{ with } \quad d=r\cdot t-s 
     \quad;\quad t=p/\pp\label{k_ddef}
\end{align}
The indices $r,s$ run over $1{\le}r{<}\pp$ and $1{\le}s{<}p$. Each conformal
weight appears twice.
For example \M 52 has $t=2/5$ and contains two independent fields 
$\phs 11=\phs 41$ and $\phs 21=\phs 31$
with conformal weights $\h 11=\h 41=0$ and
$\h 21=\h 31=-1/5$.

In a RCFT without boundaries, correlators can be expressed as a finite
sum of products of holomorphic and antiholomorphic functions, e.g. for the
4-pt-functions:
\begin{align}
     \os i \ph j(z,\bar z)\ph k(w,\bar w) \is \l
     &= \sum_{a,b}^{N<\infty} c_{ab} f_a(z,w) \bar f_b(\bar z,\bar w)
\end{align}

The functions $f_k(z,w)$ have to
obey a certain set of linear differential equations resulting from
zero norm states in the Hilbert space. Let $V_{i\l}^{jk}$ denote the
vector space of all such functions. 

We will now choose two different bases in $V_{i\l}^{jk}$.
The first one is associated with
the asymptotic behaviour of $f(z,w)$ as $w\rightarrow 0$. Following
\cite{MSb90} we introduce a pictorial notation:
\begin{align}
     \bL{\cbB ijk{\l}p{(z)}{(w)}} := f(z,w) &
     \qquad \textrm{if for $w\rightarrow 0$: } \quad
     f(1,w) = w^{h_p-h_k-h_\l}(1+\cdots) \label{g_block1}
\end{align}
The second basis is given by the behaviour as $w\rightarrow z$:
\begin{align}
     \bL{\cbC ijk{\l}q{(w)}{(z-w)}} := g(z,w) &
     \qquad \textrm{if for $w\rightarrow 1$: } \quad
     g(1,w) = (1-w)^{h_q-h_j-h_k}(1+\cdots) \label{g_block2}
\end{align}

These are just the conformal blocks as given e.g. in 
\cite{MSb90}\footnote{
  In \cite{MSb90} conformal blocks are defined in a more general
  setting using vertex operators.}, with
a particular choice of normalisation. The number of independent 3-pt
couplings of representation $i,j,k$ is given by the Verlinde fusion
numbers \N ijk which are either $0$ or $1$ for Virasoro minimal models. The
number of blocks of type \eqref{g_block1}, \eqref{g_block2} are 
$\N ijp\cdot \N klp$ and $\N jkq\cdot\N ilq$ respectively and hence are also
either $0$ or $1$.

The fusing matrix F is defined as a transformation between
the two different bases of conformal blocks:\\[-30pt]
\begin{align}
     \bL{\cbB ijk{\l}p{(z)}{(w)}} &= \sum_q \F ijk{\l}pq
     \bL{\cbC ijk{\l}q{(w)}{(z-w)}}\label{g_trans}
\end{align}
From the explicit form of the 3-point function one can check that the
fusion matrix is equal to $1$ if any of the $i,j,k,\l$ is the identity
and the resulting 3-point function is nonzero.


\section{A Recursion Formula for F-Matrices}
\label{k_section}

The F-matrices depend on the normalisation of the conformal
blocks. In the present normalisation the complicated
expressions in terms of gamma functions are hidden inside the F-matrix
entries and the expressions for the structure constants are in turn
very easy. In another normalisation the F-matrices are just the quantum
6j-symbols (see e.g. \cite{AGS89,FGP90}) and it is in principle
possible to relate the two normalisations by using expressions given
in \cite{DF84}. In the present paper a different approach has been
taken, that is the F-matrices are computed recursively. The recursion
formula given below has the advantages that it
is easy to derive, doesn't cause any problems for rational
values of the central charge, is manifestly real and doesn't introduce
sign problems through square roots (for another recursion relation
derived in a similar way see \cite{FFK89}).

To obtain the starting point of the recursion one can solve explicitly
the differential equation for the 4-pt-function
$\os i \ph j(1) \phs 21(x) \is \l$ (see e.g. \cite{YBk,DF84}).
This results in a sum involving hypergeometric functions from which
the F-matrix can be read off. The F-matrix for the (1,2) field can be
obtained from the isomorphism between \M{\pp}{p} and \M{p}{\pp}, which
amounts to the replacement $d\rightarrow -t^{-1}d$.

In the following formula $\D$ stands for either $(2,1)$
or $(1,2)$. If $j$ stands for $(r,s)$ then $j\pm\D$ is
$(r\pm 1,s)$ and $(r,s\pm 1)$ respectively. Altogether one gets:
\begin{align}
     \F ij\D\l{}{} = \begin{pmatrix}
     \textsf{F}_{\l-\D,j-\D} & \textsf{F}_{\l-\D,j+\D}\\
     \textsf{F}_{\l+\D,j-\D} & \textsf{F}_{\l+\D,j+\D}\\
     \end{pmatrix}=\cdots\notag
\end{align}
for \D=(2,1):
{\scriptsize
\begin{align}
     = \begin{pmatrix}
          \frac{\G{ d_j} \; \G{1\!-\!d_\l} }
               { \G{\tfrac{1}{2}(1\!-\!d_i\!+\!d_j\!-\!d_\l)} \; 
               \G{\tfrac{1}{2}(1\!+\!d_i\!+\!d_j\!-\!d_\l)} } & 
          \frac{\G{-d_j} \; \G{1\!-\!d_\l} }
               { \G{\tfrac{1}{2}(1\!-\!d_i\!-\!d_j\!-\!d_\l)} \; 
               \G{\tfrac{1}{2}(1\!+\!d_i\!-\!d_j\!-\!d_\l)} } \\[10pt]
          \frac{\G{ d_j} \; \G{1\!+\!d_\l} }
               { \G{\tfrac{1}{2}(1\!-\!d_i\!+\!d_j\!+\!d_\l)} \;
               \G{\tfrac{1}{2}(1\!+\!d_i\!+\!d_j\!+\!d_\l)} } & 
          \frac{\G{-d_j} \; \G{1\!+\!d_\l} }
               { \G{\tfrac{1}{2}(1\!-\!d_i\!-\!d_j\!+\!d_\l)} \;
               \G{\tfrac{1}{2}(1\!+\!d_i\!-\!d_j\!+\!d_\l)} } \\
     \end{pmatrix}\notag
\end{align}}
for \D=(1,2):
{\scriptsize
\begin{align}
     = \begin{pmatrix}
          \frac{\G{-\tfrac{1}{t}d_j} \; \G{\tfrac{1}{t}(t\!+\!d_\l)} }
               { \G{\tfrac{1}{2t}(t\!+\!d_i\!-\!d_j\!+\!d_\l)} \;
               \G{\tfrac{1}{2t}(t\!-\!d_i\!-\!d_j\!+\!d_\l)} } & 
          \frac{\G{\tfrac{1}{t} d_j} \; \G{\tfrac{1}{t}(t\!+\!d_\l)} }
               { \G{\tfrac{1}{2t}(t\!+\!d_i\!+\!d_j\!+\!d_\l)} \;
               \G{\tfrac{1}{2t}(t\!-\!d_i\!+\!d_j\!+\!d_\l)} } \\[10pt]
          \frac{\G{-\tfrac{1}{t}d_j} \; \G{\tfrac{1}{t}(t\!-\!d_\l)} }
               { \G{\tfrac{1}{2t}(t\!+\!d_i\!-\!d_j\!-\!d_\l)} \;
               \G{\tfrac{1}{2t}(t\!-\!d_i\!-\!d_j\!-\!d_\l)} } & 
          \frac{\G{\tfrac{1}{t} d_j} \; \G{\tfrac{1}{t}(t\!-\!d_\l)} }
               { \G{\tfrac{1}{2t}(t\!+\!d_i\!+\!d_j\!-\!d_\l)} \;
               \G{\tfrac{1}{2t}(t\!-\!d_i\!+\!d_j\!-\!d_\l)} } \\
     \end{pmatrix}\notag
\end{align}}
where $d_i$, $d_j$ and $d_\l$ are defined as in
\eqref{k_ddef}. Depending on the fusion rules between $i,j,\D,\l$ none
or only one of the above matrix elements may be allowed.

Using the techniques introduced in \cite{MSb90} on 5-point-functions,
or by deriving it directly from the pentagon identity,
one can obtain a recursive formula for the F-matrices:
\begin{align}
     \F ij{k+\D}\l pq &= \sum_{r,s} \F p\l\D k{k+\D,}r \cdot \F ijkrps 
     \cdot \F is\D \l rq \cdot \F q\D kjs{,k+\D} \label{k_Frec}
\end{align}
The recursion runs on the index $k$ of $\mathsf{F}_{pq}[i\l;jk]$. In
each step \eqref{k_Frec} gives the F-matrix elements for a fixed $k$ and
arbitrary (allowed) $i,j,\l,p,q$. Knowing all these on can proceed to
$k\rightarrow k+\D$.

The index range of the sum in \eqref{k_Frec} is determined
through the requirement that the eight diagrams associated to the four
F-matrices have to exist. The independent conditions on $r,s$ are:
$\N pkr\cdot\N l\D r\neq 0$, $\N jks\cdot \N q\D s\neq 0$ and
$\N rsi\neq 0$.

In particular the maximal range of the summation indices is $r=\l\pm\D$
and $s=q\pm\D$. Depending on $i,j,k,p$ it may however be smaller than
that.


\section{Boundary Conditions and Field Content}
\label{n_section}

A boundary conformal field theory (BCFT) on the upper half plane (UHP)
is specified by the field content in
the bulk $\ph i$, the possible boundary conditions which preserve
conformal symmetry $a$ and the boundary fields $\bp abi$ that interpolate
boundary conditions $a$ and $b$, as well as the structure
constants in OPEs of all these fields. Note that the boundary
conditions $a,b$ to the left and right of $x$ in $\bp abi(x)$ may or may not
be different and one can interpret the field as a boundary changing
field or a field that lives on a certain boundary, respectively.

On the UHP there is always the possibility of a boundary field inserted at
infinity. If the boundary conditions towards left and right infinity
are different this has to be the case, and if they equal there may or
may not be a field at infinity. The situation is clearer when we map
the UHP to the unit disc. Any insertion at infinity in the UHP
will appear as an additional boundary field on the unit circle. From
hereon we will only consider situations in the UHP which, when mapped
to the disc via some M\"obius transformation $z\mapsto\varphi(z)$,
have no boundary field at 
$\varphi(\infty)$. In particular on the UHP this implies that we have
same boundary 
condition towards $\pm\infty$ and no boundary field inserted at
$\infty$. This is not a restriction, because the situation just
described can be related to the general case on the UHP, i.e.\ the case with
fields inserted at infinity and possibly different boundary
conditions towards left and right infinity, via transforming it to
the unit disc, rotating the disc and mapping it back to the UHP.

There are three different OPEs to consider: Two bulk fields coming
together, expanding a bulk field in terms of boundary fields, and two
boundary fields coming together. When the bulk fields are spinless
these are, in turn\footnote{
  Note that there is a slight difference to the notation used by
  Lewellen in \cite{Lew92}. There the boundary OPE is
  $\bp abi(x) \bp bcj(y) \sim \sum_k \bc abcijk \bp ack(x)\cdot
  (y-x)^{h_k-h_i-h_j}$. The difference is in the ordering of the
  boundary fields in the n-point-functions. For the 3-pt-function the
  precise correspondence is, with $x_1>x_2>x_3$: 
  $\npt{\bp abi(x_1)\bp bcj(x_2)\bp cak(x_3)}_{\mathrm{here}} = 
  \npt{\bp ack(x_3)\bp cbj(x_2)\bp bai(x_1)}_{\mathrm{Lew.}}$, so that
  the relation between structure constants is
  $\bc abcijk(\mathrm{here})=\bc cbajik(\mathrm{Lew.})$.
  But that doesn't actually matter, because it will turn out that
  the explicit form given later fulfils $\bc abcijk=\bc cbajik$.
}:
\begin{align}
     \ph i(z)\ph j(w) &\sim
     \sum_k \C ijk \cdot \ph k(w)\cdot 
     |z-w|^{2(h_k-h_i-h_j)}\label{n_Crescale}\\
     \ph i(x+iy) &\sim
     \sum_k \B aik \cdot \bp aak(x)\cdot 
     (2y)^{h_k-2h_i}\label{n_Brescale}\\ 
     \bp abi(x) \bp bcj(y) &\sim 
     \sum_k \bc abcijk\cdot  \bp ack(y) \cdot 
     (x-y)^{h_k-h_i-h_j}\qquad ;\;x>y\label{n_bcrescale}
\end{align}

The boundary conditions of diagonal modular invariants are
given in \cite{CaL91}. A more general classification including
non-diagonal theories can be found e.g.\ in \cite{PSS96,FuS97,BPZ98}. Here
we collect some of their results we will need in the following:

For a diagonal theory the possible boundary conditions are labelled by
the bulk fields. The partition function on a cylinder of length $R$
and circumfence $L$ with boundary
conditions $a$ and $b$ is given by:
\begin{align}
   Z_{(ab)} &= \sum_i {n_{ia}}^b \cdot \chi_i(q) 
   \qquad ;\; q=e^{-\pi \frac LR}
\end{align}

In a diagonal theory the ${n_{ia}}^b$ are just the Verlinde fusion
numbers ${n_{ia}}^b=\N abi$. In particular only the identity field
lives on the $1$-boundary: ${n_{i1}}^1=\delta_{i1}$, and only the
field $\bp 1aa$ can interpolate the $1$-- and $a$-boundary
condition: ${n_{i1}}^a=\delta_{ia}$. Furthermore all bulk fields
couple to the identity on the $1$-boundary: $\B 1i1\neq 0\;\forall i$.

\section{Sewing Constraints for Diagonal Models}
\label{d_section}

In \cite{Lew92} Lewellen gives a complete set of consistency conditions for the
structure constants of a Virasoro minimal model on the UHP\footnote{
  The results in \cite{PSS96} differ from \cite{Lew92}. The equations
  given here were derived using methods from \cite{MSb90} to transform
  between the different bases of conformal blocks in the case where
  only diagonal fields are present and agree with \cite{Lew92}. We
  hope to clarify the relation to \cite{PSS96} in future work.
}. 
In the diagonal case one can, up to rescaling, determine the necessary
form of a solution to these conditions by only considering the subset
of these conditions, as given below ($1$ denotes the identity field): 
{\allowdisplaybreaks
\begin{align}
\notag\\[-22pt]
\intertext{For four boundary fields $i,j,k,\l$:}\notag\\*[-20pt]
     \raisebox{-3mm}{\figal} &= 
     \raisebox{-4mm}{\figar}\notag\\*
     \bc bcdjkq \bc abdiq\l \bc ada\l\l1 &= 
     \sum_p \bc abcijp \bc cdak\l p \bc acapp1 \cdot 
     \F ijk\l pq \label{d_bbbb}\\
\intertext{For one bulk field $i$ and two 
           boundary fields $p,q$:}\notag\\*[-20pt]
     \raisebox{-3mm}{\figbl} &= 
     \raisebox{-3mm}{\figbr}\notag\\*
     \B bi\l \bc abbp\l q \bc abaqq1 &=
     \sum_{k,m} \B aik \bc abapqk \bc aaakk1 \notag\\*
     &\qquad\quad \cdot 
      e^{i\pi(2 h_m+\frac{1}{2}h_k-h_p-h_q-2h_i+\frac{1}{2}h_\l)}\notag\\*
     &\qquad\qquad \cdot\F pqiikm \F piiqm\l\label{d_Bbb}\\[-5pt]
\intertext{For two bulk fields $i,j$:}\notag\\*[-40pt]
     \raisebox{-9.5mm}{\figcl} &= 
     \raisebox{-5.5mm}{\figcr}\notag\\*
     \B ai\l \B aj\l \bc aaa\l\l1 &=
     \sum_m \C ijm \B am1 \cdot \F jiijm\l\label{d_BB}
\end{align}}

The remaining two sewing constraints, for completeness, are:
{\allowdisplaybreaks
\begin{align}
\intertext{For two bulk fields and one boundary field:}\notag\\*[-20pt]
     \B akq \B a\l t \bc aaaqti &=
     e^{i\frac{\pi}{2}(h_t-h_i-h_q-2h_\l)}
     \sum_{p,r} \C k\l p \B api 
     e^{i\pi h_r}\notag\\*
       &\qquad \cdot
        \F pi\l kpr \F k\l rkpq \F q\l\l irt \label{d_BBb}\\
\intertext{For four bulk fields:}\notag\\*[-20pt]
     \C ijs \C k\l s \C ss1 \F ijk\l st &=
     \C jkt \C \l it \C tt1 \F i\l kjts\label{d_BBBB}
\end{align}}

These equations can be verified e.g. with the techniques for conformal
blocks described in \cite{MSb90}. With the identities on F-Matrices
given there it is also possible to see that, for real F,
the RHS of \eqref{d_Bbb} 
and \eqref{d_BBb} are equal to their complex conjugates and thus real.

Note that in the pictures associated to equations
\eqref{d_bbbb}--\eqref{d_BB} the 1-pt-function of the identity field
\bnpt 1a appears on the LHS and RHS. In the given form of the sewing
constraints these have been cancelled from the equations itself, so
that the structure constants computed from 
\eqref{d_bbbb}--\eqref{d_BB} do not depend on the 1-pt-functions of the
$1$--field. 

We still have the freedom to rescale all the
fields. This freedom must be reflected in \eqref{d_bbbb}--\eqref{d_BBBB}
in that a rescaling of the fields maps a solution to a solution.
When rescaling particular fields inside the n-pt-functions the
structure constants have to change accordingly so that the
expansions \eqref{n_Crescale}--\eqref{n_bcrescale} remain valid.
One can now verify that all possible rescalings map solutions of
\eqref{d_bbbb}--\eqref{d_BBBB} to solutions.


\section{Bulk--Boundary Couplings}
\label{p_section}

First we use the sewing constraints to fix the necessary form of the
bulk--boundary couplings. The bulk fields can be rescaled
$\ph i \rightarrow \lambda\cdot\ph i$ s.t. $\B 1i1=\S i1/\S 11$
(recall that from \cite{CaL91} we know that $\B 1i1\neq 0\;\forall
i$). Now consider \eqref{d_BB} in the form:
\begin{align}
     \sum_\l \B ai\l \B aj\l \bc aaa\l\l1 \F jjii\l n &= \C ijn \B an1 
     \label{n_alg}
\end{align}
Since by assumption only $\bp 111$ can exist on the $1$-boundary, for
$a=1$ the sum reduces to one term $\l=1$. Using \eqref{m_FFSSSS} we
obtain: 
\begin{align}
     \C ijk &= \frac{\B 1i1 \B 1j1}{\B 1k1} \F iijj1k
     = \left( \F jiijk1 \right)^{-1} \label{n_Cijk}
\end{align}

Substituting \eqref{n_Cijk} back into \eqref{d_BB} and taking
$\l=1$\footnote{proper behaviour of the identity ensures that the
  appearing boundary structure constant is equal to one (see also
  \eqref{f_ctriv} later on).} we
recover the classifying algebra for boundary conditions (i.e. the
Pasquier Algebra) \cite{PSS96,FuS97,BPZ98}, in the diagonal case:
\begin{align}
     \B ai1 \B aj1 &= \sum_k \N ijk \B ak1
\end{align}

The general solution is given by the well known eigenvalues of the fusion
matrices $\N i{}{}$. Another way to obtain the same solution is to
first find the 
general expression for the bulk--boundary couplings by
considering \eqref{d_Bbb} with $a=1, p=q=b$ (the $k$--sum reduces to $k=1$):
\begin{align}
     \B bi\l \bc 1bbb\l b &= \B 1i1
     \sum_{m} e^{i\pi(2(h_m-h_b-h_i)+\frac{1}{2}h_\l)}
     \F bbii1m \F biibm\l\label{n_Bbb}
\end{align}
The constant \B 1i1 is fixed by the normalisation of the bulk fields
and the imaginary part of the sum can be shown to vanish, so that 
RHS is known and real. 
Petkova has observed \cite{VaP98} that up to a normalisation the
$\B aik$ are just the S-matrix elements $\S ai(k)$ for the torus with one
operator insertion\footnote{
  see also \cite{BPPZXX} for a general discussion of the
  relation between the sets of duality relations in \cite{Lew92} and
  \cite{MSb90}}.
This S-Matrix is given by equation 
\eqref{m_S} in the appendix. For $k{=}1$ this reduces to the S-matrix
that implements the modular 
transformation of characters: $\S ij(1) = \S ij$.

Taking \eqref{n_Bbb} and rearranging terms such that a rescaling invariant
combination of structure constants appears on the LHS, together with
\eqref{m_FFFF}--\eqref{m_S} the correspondence
between $\B aik$ and $\S ai(k)$ becomes: 
\begin{align}
     \frac{\B aik \bc 1aaaka}{\B 1i1} 
     &= e^{i\frac\pi2 h_k}
     \left(\F kaaka1\right)^{-1} 
     \cdot \frac{\S 11\cdot \S ai(k)}{\S 1i\cdot\S a1}
     \label{n_B}
\end{align}
This expression is symmetric under the exchange 
$a\leftrightarrow i$ and, since the phase on the RHS cancels with the
phases in $\S ai(k)$, it is also real. Using the explicit normalisations
$\B 1i1{=}\S i1/\S 11$ and $\bc 1aaaka{=}1$ (see equation \eqref{f_bsc}
later on) \eqref{n_B} simplifies to 
\begin{align}
     \B aik &= e^{i\frac\pi2 h_k}
     \left(\F kaaka1\right)^{-1} 
     \cdot \frac{\S ai(k)}{\S a1}\qquad\qquad
     \B ai1 = \frac{\S ai}{\S a1}\label{n_Bbi1}
\end{align}

All the bulk and bulk--boundary structure constants are now determined
and the scaling of all bulk fields is fixed.


\section{Boundary Structure Constants}
\label{f_section}

To give necessary expressions for the boundary structure constants it
is enough only to consider equation \eqref{d_bbbb}. The following
table summarises the line of argument in this section by describing
the boundary situations considered, which constant it fixes, and what
the remaining freedom to rescale the boundary fields is:\\[3pt]

\centerline{\begin{tabular}{|l|l|l|l|}
\hline
&&&\\[-10pt]
boundary situation & s.c. fixed & eqn. & remaining freedom \\[1mm]
\hline
&&&\\[-10pt]
\ffbnd 1bcbbjjb & \bc bcbjj1 & \eqref{f_Cii1} &
     $b{\neq} c, j{\neq} 1 : \{\bp bcj,\bp cbj\} \rightarrow$ \\
&&&\hspace{2cm} $\{\lambda\cdot\bp bcj,\lambda^{-1}\cdot\bp cbj\}$ \\
&&&   $b = c, j{\neq} 1 : \{\bp bbj\} \rightarrow \{\pm\bp bbj\}$ \\
\ffbnd 1b1dbbdd & \bc b1dbdq & \eqref{f_Cbdq} & none \\[2mm]
\ffbnd ab1dibd\l & \bc dab\l iq & \eqref{f_bsc} &  none \\[2mm]
\hline
\end{tabular}}
~\\[2pt]
In order to simplify the notation when rescaling subsets of boundary
operators it is helpful to introduce an ordering on the boundary
conditions $(1)<(a_1)<(a_2)<\dots$. The particular order one chooses
is not important.

As a first step consider the relation coming from taking the two
different OPE's in the 3-point function:
\begin{align}
     \bc abcijk \bc acakk1 &= \bc bcajki \bc abaii1 \label{f_bbb}
\end{align}

Setting $i{=}1, k{=}j, b{=}a$ in \eqref{f_bbb} leads to
$\bc aac1jj {=} \bc aaa111$. Similarly, setting $j{=}1$ resp. $k{=}1$ leads
to $\bc abbi1i {=} \bc bba1ii$ resp. $\bc aaa111 {=} \bc baa11i$. It
follows that consistent behaviour of the identity field on the
1-boundary $\bc 111111{=}1$ already implies consistent behaviour of the
identity on all other boundaries:
\begin{align}
     \bc aab1ii = 1 \qquad \bc abbi1i = 1 
     \qquad \forall a,b,i \label{f_ctriv}
\end{align}
Any given solution of the sewing constraints
\eqref{d_bbbb}--\eqref{d_BBBB} must already fulfill \eqref{f_ctriv}.

To see how boundary operators couple to the identity, first consider
\eqref{d_bbbb} with $a{=}1$, $d{=}b$, $i{=}\l{=}b$, $j{=}k$, $q{=}1$:
The sum reduces to 
$p{=}c$ and with \eqref{f_bbb} in the form 
$\bc cb1jbc \bc 1c1cc1 {=} \bc 1cbcjb \bc 1b1bb1$ one obtains:
\begin{align}
     \bc bcbjj1 \left(\F bjjbc1\right)^{-1} &= 
     \bc 1bcbjc \bc 1cbcjb
\end{align}
Exchanging $b \leftrightarrow c$ leaves the RHS invariant. Transforming
the F-matrix element using \eqref{m_FSSF} we finally get:
\begin{align}
     \S 1c \bc cbcjj1 = \S 1b \bc bcbjj1 \label{f_ScSc}
\end{align}
Any solution has to fulfill this identity. In particular this implies
that independent of the expectation values of the identity $\bnpt 1a$
we cannot set both $\bc cbcjj1$ and $\bc bcbjj1$ to one.

We shall now make use of the freedom to rescale the fields. Note that
the identity field $1$ is fixed by the property $1\cdot 1{=}1$. For
$b<c$ and $j{\neq} 1$ we rescale
$\bp bcj \rightarrow \lambda\cdot\bp bcj$. By \eqref{n_bcrescale}
this results in:
\begin{align}
     \bc cbcjj1 \rightarrow \lambda \bc cbcjj1
     \qquad
     \bc bcbjj1 \rightarrow \lambda \bc bcbjj1\label{f_Cresc}
\end{align}

For an appropriate $\lambda$ we get, for all $b<c, j{\neq}1$: 
\begin{align}
     \bc cbcjj1 &= \left( \F ccbb1j \right)^{-1} \frac{\S 11}{\S 1c}
     \label{f_Cii1}
\end{align}
In particular \eqref{f_Cresc} implies that once the $\bc cbcjj1$ are
adjusted for $b<c$ one is no longer free to rescale $\bc cbcjj1$ for
$b>c$. However \eqref{f_ScSc} implies that \eqref{f_Cii1}
holds also for $b>c, j{\neq}1$.For $b{=}c, j{\neq}1$ rescaling
$\bp bbj \rightarrow\lambda\cdot\bp bbj$ gives
$\bc bbbjj1 \rightarrow \lambda^2 \bc bbbjj1$, so that bringing
$\bc bbbjj1$ to the form \eqref{f_Cii1} only fixes $\bp bbj$ up to a
sign. Setting $j{=}1$, together with
\eqref{m_FSS} shows consistency with $\bc bbb111 {=}1$.  Thus
\eqref{f_Cii1} is valid for all values of $b,c,j$. The 
scaling of the operators $\{\bp bcj,\bp cbj\}$ for
$b{\neq} c, j{\neq} 1$ is 
now fixed up to 
$\{\bp bcj,\bp cbj\} \rightarrow 
\{\lambda\cdot\bp bcj,\lambda^{-1}\cdot\bp cbj\}$, which leaves
\eqref{f_Cii1} invariant. $\bp bbj, j{\neq} 1$ is fixed up to sign.

Taking \eqref{d_bbbb} with $a{=}c{=}1, i{=}j{=}b, k{=}\l{=}d, p{=}1$ and using
\eqref{f_bbb} and \eqref{f_Cii1} results in:
\begin{align}
     \bc b1dbdq \bc d1bdbq &= \left(\F ddbb1q \right)^2 \label{f_ccF}
\end{align}

For $b>d, q{\neq} 1$ we rescale $\{\bp dbq,\bp bdq\} \rightarrow 
\{\lambda\cdot\bp dbq,\lambda^{-1}\cdot\bp bdq\}$ such that:
\begin{align}
     \bc b1dbdq = \F ddbb1q \label{f_Cbdq}
\end{align}
\eqref{f_ccF} now implies that \eqref{f_Cbdq} also holds for
$b<d$. For $b{=}d, q{\neq}1$ we are still free to choose the sign of
$\bp bbq \rightarrow \pm \bp bbq$. This allows us to alter the sign of  
$\bc b1bbbq$ to match \eqref{f_Cbdq}. The case $q{=}1$ can occur only
for $b{=}d$ and we get $\bc b1bbb1 {=} \S 11 / \S 1b$,
consistent with the normalisation \eqref{f_Cii1}. Thus \eqref{f_Cbdq}
holds for all $b,d,q$. The scaling of all boundary operators is now
fixed.

Taking \eqref{d_bbbb} with $c{=}1, j{=}b, k{=}d$ the sum reduces to
$p{=}a$. Using \eqref{f_bbb} and rearranging terms one obtains:
\begin{align}
     \bc dab\l iq = \F ibd\l aq
     \frac{\bc abaii1 \bc dad\l\l1}{\bc dbdqq1 \bc a1aaa1}
     \frac{\bc b1abai \bc a1dad\l}{\bc b1dbdq}
\end{align}

All the C-terms cancel and after renaming indices one is left with:
\begin{align}
     \bc abcijk &= \F iacjbk \qquad \forall a,b,c,i,j,k \label{f_bsc}
\end{align}

Equations \eqref{n_Cijk}, \eqref{n_Bbb} and \eqref{f_bsc} are the unique solution (if
it exists) to the sewing constraints. All other solutions are
trivially related by a rescaling of the fields.


\section{One Point Function of the Identity Operator}
\label{h_section}

Recall that the 1-pt-function of the 1-operator cancelled in
equation \eqref{d_bbbb}--\eqref{d_BB} (and in
\eqref{d_BBb}, \eqref{d_BBBB} as well in fact).
We can fix these by demanding
that the partition function around a cylinder of length $R$ and
circumfence $L$ is a modular
transformation of the partition function along the cylinder
\cite{Car89,BPZ98}.

The partition function around the cylinder is equivalent to a trace on the
UHP which reduces to a sum of characters:
\begin{align}
     Z_{(ab)}\left[\hhpicUHP\right] 
     &= \sum_j {n_{ja}}^b \chi_j(q)
      = \sum_{i,j} \frac{\S ai \S bi}{\S 1i} 
        \cdot \S ij \chi_j(q)
      = \sum_{i} \frac{\S ai \S bi}{\S 1i} 
        \cdot \chi_i(\tilde q) \label{h_ZUHP}
\end{align}
where $q=\exp(-\pi L/R)$ and $\tilde q=\exp(-4\pi R/L)$.

The partition function along the cylinder in turn is equivalent to an
inner product of boundary states. Let
$\ishin i=(1+(2h_i)^{-1}L_{-1}\bar L_{-1}+\cdots)\is i$ 
denote the Ishibashi
states \cite{Ish89} and $\is a = \sum_i g_{ai} \ishin i$ be a boundary
state. It will turn out that there is a consistent choice of constants
such that the $g_{ai}$ are real and complex conjugation has thus been
left out of subsequent formulas. For the partition function we get:
\begin{align}
     Z_{(ab)}\left[\hhpicRING\right]
     &= \os b e^{-\frac{2\pi R}{L}(L_0+\bar L_0-\frac{c}{12})} \is a = \sum_i 
     g_{ bi} \, g_{a i} \; \npt{i|i}
     \cdot \chi_i(\tilde q) \label{h_Zring}
\end{align}

Comparing \eqref{h_Zring} to \eqref{h_ZUHP} gives expressions for
$g_{a i}$. Alternatively these can be obtained by calculating the
1-pt-function on a disc in two different ways. 
Consider a bulk field $\ph i$ in the centre of a disc of radius 1 with
boundary condition $b$. Using Ishibashi
states one gets:
\begin{align}
     \bnpt{\ph i(0)}{b}_{\mathrm{disc}} &= \os b\ph i(0)\is 0 
     = g_{bi} \npt{i|i}
\end{align}
Here we took the states \is i to be normalised such that:
\begin{align}
     \ph i(0) \is 0 = \is i \label{h_inorm}
\end{align}

On the other hand, transforming the the 1-pt-function from the UHP with a
$b$-boundary to the disc we get:
\begin{align}
     \bnpt{\ph i(0)}{b}_{\mathrm{disc}} &=
     \B bi1 \bnpt 1b
\end{align}
Comparing the two yields:
\begin{align}
     g_{bi} = 
     \frac{\B bi1 \bnpt 1b}{\npt{i|i}} \label{h_gbi}
\end{align}

The norm of the state \is i is linked to OPE of two bulk fields via
\eqref{h_inorm}:
\begin{align}
     \npt{i|i} &= \C ii1 \npt{0|0}
\end{align}
To ensure that the boundary 1-pt-functions of the identity are real
and take a concise form we choose not to set $\npt{0|0}$ to one.

Substituting \eqref{h_gbi} back into \eqref{h_Zring} gives:
\begin{align}
     g_{ai}\; g_{bi}\; \npt{i|i} &= 
     \frac{\B ai1 \B bi1}{\C ii1} \cdot
     \frac{\bnpt 1a\bnpt 1b}{\npt{0|0}} \label{h_ggii}
\end{align}
The RHS of this equation is invariant under a rescaling
$\ph i\rightarrow\lambda\ph i$ because the factors from
$\B ai1 \cdot \B bi1$ and $\C ii1$ cancel. Thus substituting the
explicit expressions for \B bi1 \eqref{n_Bbi1} and \C ii1
\eqref{n_Cijk} still gives an expression that is invariant under
rescaling of the fields. Comparing \eqref{h_Zring} to \eqref{h_ZUHP}
now yields: 
\begin{align}
     \frac{\bnpt 1a\bnpt 1b}{\npt{0|0}}
     &= \frac{\S a1 \S b1}{\S 11} \label{h_modconst}
\end{align}

The general solution to \eqref{h_modconst} is:
\begin{align}
     \bnpt 1a = \mu\cdot \S a1
     &\qquad \npt{0|0} = \mu^2\cdot\S 11\label{h_1pt}
\end{align} 
For convenience we choose $\mu=1$.

There is another constraint on the identity 1-pt-functions, coming
from demanding continuity on the disc. We take boundary fields to
transform as
\hbox{$\bp abi(x) \rightarrow 
      |\varphi'(x)|^{h_i} \bp abi(\varphi(x))$}.
If we lift the two UHP 2-pt-functions 
$\bnpt{\bp abi(x)\bp bai(y)}a_{UHP}$ and
$\bnpt{\bp bai(y)\bp abi(x)}b_{UHP}$ to the disc, they should be
identical upon analytic continuation. This results in the constraint
(see also \cite{Lew92}):
\begin{align}
     \bc abaii1 \bnpt 1a &= \bc babii1 \bnpt 1b \label{h_1ptcont}
\end{align}
Comparing \eqref{h_1ptcont} with \eqref{f_ScSc}, independent of the
normalisation of boundary fields, this becomes:
\begin{align}
     \frac{\bnpt 1a}{\S 1a} &= \frac{\bnpt 1b}{\S 1b} \label{h_discconst}
\end{align}
The general solution for $\bnpt 1a$ in \eqref{h_1pt} is thus
consistent with demanding continuity on the disc, for any value of $\mu$.

One can now work out the universal ground state degeneracies defined
in \cite{AfL91}:
\begin{align}
     Z_{(ab)} &\simu R\infty (g_a \cdot g_b) \cdot e^{-R E_0(L)} 
\end{align}
where $E_0(L)$ is the ground state energy and $g_a$ resp.\ $g_b$ is the factor
of the ground state degeneracy coming from the boundary $a$ resp.\ $b$.
Let $\ph\O$ be the field of lowest conformal weight. Then the
S-matrix (for Virasoro minimal models) satisfies
$\S \O a>0\; \forall a$. From \eqref{h_Zring} we
get the relation $g_a\cdot g_b = g_{a\O}\cdot g_{b\O}\cdot \npt{\O|\O}$. Thus
\begin{align}
     g_a &= g_{a\O} \sqrt{\npt{\O|\O}}
     = \frac{\B a\O1\cdot\bnpt 1a}{\sqrt{\C\O\O1\npt{0|0}}}
     = \frac{\S a\O}{\sqrt{\S 1\O}}
\end{align}
in agreement with \cite{AfL91}.
Note that this expression is invariant under both rescaling of the fields
and rescaling of the identity-1-pt-functions.


\section{Conclusion}
\label{o_section}

In this paper we have found the structure constants of A-type
Virasoro minimal models, given in equations \eqref{n_Cijk},
\eqref{n_Bbb} and \eqref{f_bsc}, under the assumption that a solution
to the sewing constraints exists. Any other solution is trivially
related to the given set of structure constants by a rescaling of the
fields. Numerical tests with randomly chosen fields in several A-type
Virasoro minimal models confirm that the given structure constants
solve the sewing constraints \eqref{d_bbbb}--\eqref{d_BBBB}.

In the form given in section \ref{d_section} the sewing constraints do
not depend on the boundary 1-pt-function of the identity $\bnpt
1a$. In particular there is no normalisation of the boundary fields
s.t.\ for $a\neq b$ we can have both $\bc abaii1=1$ and $\bc babii1=1$
(see equation \eqref{f_ScSc}). Also none 
of the structure constants given depends on the final choice for the
values of $\bnpt 1a$.

There are two constraints on the $\bnpt 1a$ and the normalisation of
the bulk-vacuum $\npt{0|0}$: one from modular
transformation of the partition function \eqref{h_modconst} and one from
demanding continuity on the disc \eqref{h_discconst}. In fact these
are equivalent, because on the one hand, once $\npt{0|0}$
is chosen, \eqref{h_modconst} actually determines the $\bnpt 1a$ up to
an overall sign and one can check that \eqref{h_discconst} is
satisfied, irrespective of the choice of that sign. On the other hand,
once \eqref{h_discconst} is solved, we can choose $\npt{0|0}$ such that
\eqref{h_modconst} is also satisfied.
The choice for  $\bnpt 1a$  and  $\npt{0|0}$ is not affected by
rescaling of the bulk or boundary fields.

The individual rescalings are, for
$\alpha_i,\beta^{(ab)}_i,\mu\in\mathbb{C}$:
\begin{align}
     \ph i &\rightarrow \alpha_i \cdot \ph i \notag\\
     \bp abi &\rightarrow \beta^{(ab)}_i \cdot \bp abi \notag\\
     \npt{0|0}, \bnpt 1a,\bnpt 1b,\dots &\rightarrow
     \mu^2\npt{0|0}, \mu\bnpt 1a,\mu\bnpt 1b,\dots\notag
\end{align}
To relate the normalisation used in this paper to others one can look
at the scale dependant form of the following structure constants:
\begin{align}
     &\B ai1 = \alpha_i\cdot \frac{\S ai}{\S a1}
     &&\C ii1 = (\alpha_i)^2\cdot \frac{\S 1i}{\S 11}
     &&\npt{0|0}= \mu^2\cdot\S 11
     &&\bnpt 1a = \mu\cdot\S a1\notag
\end{align}
In this paper we have $\alpha_i=1$, $\mu=1$. To recover the
normalisation of Cardy and
Lewellen \cite{CaL91,Lew92} one has to set
$\alpha_i = \sqrt{\S 11/\S1i}$ and $\mu=1/\sqrt{\S 11}$.

The ground state degeneracies have been computed in this formalism and
the result agrees with \cite{AfL91} and is, as expected, invariant
under all rescalings.

\vspace{0.5cm}
\noindent{\bf Acknowledgements} ---
This work was supported by the EPSRC, the DAAD and King's College
London. I especially wish to thank my
supervisor G.M.T.\ Watts for suggesting the project, encouragement,
criticism and helpful discussions, as well as for sharing the fruits
of conversations with J.-B.\ Zuber at meetings in SISSA, DESY and Durham.
Furthermore I want to thank V.B.\ Petkova for several useful
conversations and I am grateful to her, A.\ Sagnotti,
Ya.S.\ Stanev and J.-B.\ Zuber for comments and corrections to early
versions of this paper. 


\section{Appendix: Some Useful Identities for F-Matrices}
\label{m_section}

The following identities from \cite{MSb90} for Virasoro minimal models
are useful when working with the structure constants:
\begin{align}
     &\F ijk\l pq = \F ji\l kpq = \F k\l ijpq \\
     &\F iiii11 = \frac{\S 11}{\S 1i}\label{m_FSS} \\
     &\F iijj1k \F jiijk1 = \frac{\S 11\cdot \S 1k}{\S 1i\cdot\S 1j} 
        \label{m_FFSSSS}\\
     &\F jiijk1 = \frac{\S 1k}{\S 1j} \cdot \F ikkij1\label{m_FSSF}\\
     &\F \l ii\l n1 \F njk\l pi = \F \l kk\l p1 \F \l ijpnk 
      \label{m_FFFF}\\
     &\Bmat \epsilon ijk\l pq = e^{i\pi\epsilon (h_i+h_\l-h_p-h_q)}
      \F ijlkpq\label{m_BF}\\
     &\S ij(p) = 
      \S 11 e^{-i\pi h_p} \frac{\F piipi1}{
          \F pppp11 \F jjjjp1 \F iiiip1}
      \sum_r \Bmat -iijjpr \Bmat -ijijr1\label{m_S}
\end{align}

\end{document}